\documentclass{tlp}
\usepackage{amsmath}
\usepackage{graphicx}
\usepackage{multirow}
\usepackage{pgfplots}
\usepackage{listings}
\usepackage{xcolor}
\usepackage{array}
\usepackage{booktabs}
\usepackage[utf8]{inputenc}
\usepackage{decision-table}
\usepackage{makecell}
\usepackage{multirow}
\usepackage{todonotes}
\usepackage{makecell}
\usepackage{blindtext}
\usepackage{xspace}
\usepackage{amssymb}
\usepackage{subcaption}
\usepackage[hidelinks]{hyperref}
\usepackage{colortab}
\usepackage{algorithm}
\usepackage[noend]{algpseudocode}
\usepackage{graphicx}
\usepackage{tcolorbox}
\usepackage{inconsolata}
\usepackage{xcolor}
\usepackage{listings}
\usepackage{caption}
\def\ContinueLineNumber{\lstset{firstnumber=last}}

\DeclareCaptionFont{white}{\color{white}}
\DeclareCaptionFormat{code}{%
  \parbox{\textwidth}{\colorbox{gray}{\parbox{\textwidth}{#1#2#3}}}}
\usepackage{titlesec}
\usepackage{eurosym}
\usepackage{enumitem}
\usepackage{amsmath}
\DeclareCaptionFormat{listing}{
\parbox[bottom][0pt][c]{\textwidth}{\hspace{0pt}#1#2#3}}
\DeclareCaptionFont{black}{\color{black}}
\pgfplotsset{compat=1.15}
\lefttitle{ASP for Flexible Payroll Management}

\jnlPage{X}{X}
\jnlDoiYr{2024}
\doival{10.1017/xxxxx}

\title[ASP for Flexible Payroll Management]
        {Multi-Shot Answer Set Programming for Flexible Payroll Management}

\begin{document}
\begin{authgrp}
\author{\sn{Benjamin} \gn{Callewaert}}
\affiliation{KU Leuven, De Nayer Campus, Dept. of Computer Science \\
J.-P. De Nayerlaan 5, 2860 Sint-Katelijne-Waver, Belgium \\
Leuven.AI - KU Leuven Institute for AI, B-3000 Leuven, Belgium}
\author{\sn{Joost} \gn{Vennekens}}
\affiliation{KU Leuven, De Nayer Campus, Dept. of Computer Science \\
J.-P. De Nayerlaan 5, 2860 Sint-Katelijne-Waver, Belgium \\
Leuven.AI - KU Leuven Institute for AI, B-3000 Leuven, Belgium}
\end{authgrp}
\maketitle
\begin{abstract}
    Payroll management is a critical business task that is subject to a large number of rules, which vary widely between companies, sectors, and countries. Moreover, the rules are often complex and change regularly. Therefore, payroll management systems must be flexible in design. In this paper, we suggest an approach based on a flexible Answer Set Programming (ASP) model and an easy-to-read tabular representation based on the Decision Model and Notation (DMN) standard. It allows HR consultants to represent complex rules without the need for a software engineer, and to ultimately design payroll systems for a variety of different scenarios. We show how the multi-shot solving capabilities of the \emph{clingo} ASP system can be used to reach the performance that is necessary to handle real-world instances. Under consideration in Theory and Practice of Logic Programming (TPLP).
\end{abstract}
Competing interests: The author(s) declare none
\section{Introduction}
\ContinueLineNumber
Payroll management is a critical business task that concerns the administration and management of staff financial reports, such as wages, salaries, deductions, and bonuses. 
Manual preparation of staff’s salaries is often error-prone and time-consuming due to the large number of relevant rules. Automated payroll management systems are therefore being used to speed up the process. 
The set of applicable rules can vary widely based on the sector and the country in which the company operates, and on company-specific agreements that have been made.
Correctly implementing and maintaining a payroll system can thus be a challenging exercise \citep{doody1982report}. 
Based on talks with the company ProTime, a market leader in the area of time registration, we have identified the following three key challenges for such a system.

\begin{enumerate}
    \item When deploying the system for a new company or updating it for an existing customer, HR consultants are typically employed to figure out the rules that apply. If they need to communicate all their  knowledge 
    to software engineers before it can be entered into the system, this introduces a lot of overhead, delays, and the risk of communication errors. Therefore, the HR consultants should be able to configure as much of the system as possible without the help of a software engineer. Due to the complexity of the rules, a simple configuration file does not suffice and a more elaborate knowledge representation language is needed. Providing such a language that is both powerful enough and easy to use for HR consultants is an important challenge.
    \item There are essentially no restrictions on the kinds of rules that the HR consultants may encounter. It is therefore not possible to cover all the expressivity that the HR consultants might need up-front. Therefore, the language in which they write down the rules must not only be easy to use for them, but it should also be possible to easily extend it with new language features, without invalidating models that have been built earlier.
    \item Finally, despite the required flexibility, the solution should still be computationally efficient. In particular, a single employee shift should be handled in $< 1s$. This is very challenging, as shifts may run over several days and it is necessary to determine the employee's wages at each point in time. Moreover, it is impossible to know up-front at which points in time the pay rate will change, because this can be determined by the rules in a complex way.
\end{enumerate}
In this paper, we propose an approach based on a combination of multi-shot Answer Set Programming and decision tables to tackle these challenges. Throughout the paper, we use the following real-life example to illustrate our approach:

\begin{quote}
   An employee receives a normal wage of 20\euro\ per hour and an overtime premium of 20\% for any work done after 8 hours of working. The employee should also receive a night premium of 25\% for any work done at night or any work done in the evening that continues into the night, provided that more time was spent in the night than in the evening. Employees are also allowed to take a break, but breaks are only paid as (official) rest breaks after their shift has been going for at least one hour.  
\end{quote}
The paper is structured as follows: we begin by introducing Answer Set Programming, the Decision Model and Notation standard, and Event Calculus in Section~\ref{Preliminaries}.
Next, we introduce a discrete, multi-valued variant of the Event Calculus in Section~\ref{s:DFEC}. After that, we elaborate our approach to tackle the above-mentioned key challenges of a payroll system in Section~\ref{proposedapproach} and present a first implementation of it in Section~\ref{ssimplementation}. We present a more efficient multi-shot implementation in Section~\ref{msimplementation}. After that, we discuss the implementations and their results in Section~\ref{benchmark} and discuss related work in Section~\ref{relatedwork}.
Finally, we conclude in Section~\ref{conclusion}.
\section{Preliminaries}\label{Preliminaries}
\subsection{Answer Set Programming}
Answer Set Programming (ASP) is a declarative paradigm for solving knowledge-intensive computational problems \citep{brewka2011answer,gebser2012answer}. The idea of ASP is to represent a problem as a logic program whose models, called \textbf{answer sets} \citep{gelfond1991classical} correspond to the solutions. A normal ASP logic program $\Pi$ is a finite set of rules $r$ of the form:
\begin{equation*}
    a_{1} :- a_{m+1}, \ldots, a_{n}, not \ a_{n+1}, \ldots, not \ a_{l}.
\end{equation*}
where each $a_{i}$ is an atom of the form $p(t_{1},..., t_{k})$ with $p$ being a predicate symbol and $t_{1},\ldots, t_{k}$ terms. Atom $a_{1}$ is called the head atom, $head(r)$, while $a_{m+1}$ to $a_{n}$ and $not\ a_{n+1}$ to $not\ a_{l}$ are referred to as positive and negative body literals. 
The positive and negative sets of atoms ($body^+(r)$ and $body^-(r)$ respectively) compose the body of rule r, referred to as $body(r)$. 
The informal meaning of a normal rule is that if the body is true, then so is the head.
A rule $r$ is called a \textbf{fact} if it consists of only one head atom and no body atoms.
An \textbf{integrity constraint} $r$ is a rule that contains no head atom. An integrity constraint $r$ is satisfied if $body(r)$ is false. By insisting that integrity constraints are satisfied in an answer set, they effectively rule out answer sets for which this is not the case. 
An expression is said to be ground if it contains no variables. Furthermore, in the original (non-ground) program, every normal rule or constraint $r$ must be safe; that is, every variable which occurs in $r$ must occur at least once in $body^+(r)$. The logic operator $\mathit{not}$ denotes negation as failure, allowing ASP to perform non-monotonic reasoning \citep{nafkeith1977}. 
\textbf{Aggregate functions}, denoted as $f(S)$, where $S$ is a set and $f$ can be $\#count$, $\#min$, $\#max$, $\#sum$, or $\#times$, are represented as $L_g \prec_1 f(S) \prec_2 R_g$. Here, $f(S)$ is the aggregate function, $\prec_1, \prec_2 \in \{=, <, \leq, >, \geq\}$, and $L_g$ (left guard) and $R_g$(right guard) are terms. Either $L_g \prec_1$ or $\prec_2 R_g$ can be omitted, defaulting to "0" and "$+\infty$" respectively \citep{faber2008design}. 
The standard semantics of aggregates include counting, finding the minimum and maximum, summing, and multiplying atoms in a multiset $S$.

Semantically, a logic program induces a set of stable models, being distinguished models of the program determined by the stable model semantics \citep{gelfond1988stable}. ASP systems like Clingo \citep{gebser2014clingo} and DLV \citep{leone2006dlv} use a grounder to replace variables with constants and a solver to search for answer sets. 
\subsubsection{Multi-shot Answer Set Programming}
Standard ASP always follows this two-step process in computing the answer sets of logic programs. However, this paradigm is limited in the sense that it cannot handle efficiently cases in which data or constraints should be added, deleted, or replaced as part of the solving process.

Recently,~\citeN{gebser2019multi} extended the traditional ground-and-solve paradigm of ASP to \textbf{multi-shot solving}. Here, the user defines a number of parametrised ASP programs and writes an imperative program, e.g. in Python, to manipulate these programs. The idea is to consider evolving grounding and solving processes where the internal state of the logic program can be manipulated by certain operations. Such operations allow for adding, grounding, and solving logic programs as well as setting truth values of (external) atoms~\citep{gebser2019multi}. \textbf{Multi-shot ASP} solving is an iterative approach that incorporates this idea and is geared for problems where the logic program is continuously changing. To this end, \emph{clingo} complements ASP’s declarative input language with different ways of controlling the grounding and solving process. An imperative programming interface allows a continuous assembly of the program and gives control over the grounding and solving functions. A new \#program directive is also introduced that allows to structure a program into subprograms or modules, making the solving process completely modular. Each subprogram has a name and an optional list of parameters. Subprogram \emph{base} is a dedicated subprogram that collects all the rules not preceded by any \#program directive. Finally, \#external directives are used within subprograms to set external atoms to some truth value via the Python interface of clingo. Rules that include such external atoms in the body can be selectively (de)activated to direct the search.

The \textbf{Module Theorem} \citep{ModuleTheorem2006,ModuleTheorem2009}, lays out a theoretical foundation for establishing a modular structure in logic programs operating under the stable model semantics. This theoretical framework ensures that each module possesses a clearly defined input/output interface, crucial for achieving composability among answer sets of distinct modules. As long as the requirements outlined in the Module Theorem are met, we can carry out the module-by-module computation of answer sets for an ASP program. 
Two critical implications arise from the concept of compositionality as defined in the Module Theorem. Firstly, it mandates that all rules defining a specific (ground) literal must be contained within the same module. Secondly, the Module Theorem specifies that positive rule cycles must be confined within individual modules. This theoretical requirement is essential for preventing unintended interactions between modules and preserving the integrity of the computational process.

\subsection{DMN}
Decision Model and Notation (DMN) \citep{OMGspec} is a standardized modelling language and notation for expressing business rules. It is designed to be easy to use and understand by business experts \citep{silver2016dmn}, while still being expressive enough to capture complex decision-making logic. DMN models consist of two main components: a Decision Requirements Diagram (DRD) and one or more decision tables.

The DRD is a graphical representation of the overall structure and dependencies of a DMN model. It shows the connections between the various inputs, decisions, and knowledge sources that are used in the model, making it easier for end-users to interpret and understand the model \citep{hasic2017}.

\begin{figure}
    \centering
    \scriptsize
    \dmntable{Employee Bonus}{U}{Employee's Role, Years Worked}{Receives Bonus}
             {Manager, ---, Yes,
              Employee, $\leq 1$, No,
              Employee, $> 1$, Yes,
              Worker, $\leq 3$, No,
              Worker, $> 3$, Yes}
    \caption{Example of DMN table determining eligibility for holiday bonus}
    \label{fig:DMN_EmployeeBonus}
\end{figure}

Decision tables contain the detailed business logic that is used to make decisions. Each decision table consists of a number of input columns and at least one output column. Rows in the table specify the conditions under which a given output should be produced, based on the values of the input columns. A classical example of a decision table is shown in Figure \ref{fig:DMN_EmployeeBonus}. It defines whether an employee receives a holiday bonus based on their role in the company and the number of years they have worked at the company. The decision table has input columns for the employee's role and the number of years they have worked, and an output column for whether they receive a bonus. Rows in the table correspond to different kinds of employees and specify for each kind whether they are eligible for a bonus. For example, a manager always receives a holiday bonus while a worker only receives one if they have worked for more than 3 years at the company.

Overall, DMN provides a user-friendly and intuitive way of expressing business rules and decision-making logic, making it well-suited for use in a variety of applications \citep{sooter2019modeling,hasic2020decision}. It has already been successfully used in many case studies and is becoming increasingly popular as a modelling language for decision management \citep{hasic2016}.

\subsection{Event Calculus} 
The Event Calculus (EC) is a logical formalism used in artificial intelligence and knowledge representation to reason about events and their effects \citep{kowalski1986logic}. It is based on first-order logic and has a temporal component, allowing it to represent the changing state of a system over time. The Event Calculus has been applied to a wide range of AI tasks, including planning \citep{eshghi1988abductive,shanahan2000abductive}, cognitive robotics \citep{shanahan1996robotics,shanahan1999ramification} and legal reasoning \citep{kowalski1995legislation}. It has also been used to model the effects of actions in multi-agent systems, allowing for the formalization and analysis of complex interactions between agents \citep{kowalski1999logic}. A complete EC model consists both of domain-independent and domain-dependent axioms. The general theory of how properties evolve through time is captured in the domain-independent axioms, which cover notions such as persistence and causality. The domain-dependent axioms, on the other hand, describe which actions affect which fluents under various circumstances in a certain domain, and state when these actions do occur.

The original formulation of Event Calculus uses circumscription \citep{mccarthy1980circumscription}, or the minimization of the extensions of predicates, to allow default reasoning. In particular, it reflects the assumptions that (1) the only events that occur are those known to occur and (2) the only effects of events are those that are known. The closed world assumption of most Logic Programming semantics, including the stable model semantics, allows the same assumptions to be incorporated in an arguably more elegant way. This makes Answer Set Programming a very suitable formalism for implementing the EC. 

Different variants of the Event Calculus have been proposed over the years; in this paper, we will use an implementation based on the Functional Event Calculus (FEC) \citep{ma2014epistemic} and Discrete Event Calculus (DEC) \citep{DEC}.

The \textbf{FEC} extends the Event Calculus with non-boolean fluents. This sets the EC on a par with formalisms such as the Situation Calculus in this respect. In the FEC, there are four sorts: $A$ for actions ($a, a', a_{1},\ldots$), $F$ for fluents ($f, f', f_{1},\ldots$), $V$ for values ($v, v', v_{1},\ldots$), and $T$ for timepoints ($t, t', t_{1},\ldots$).
The key predicates and functions of the FEC are:
$\mathit{Happens} \subseteq A \times T$,
$\mathit{ValueOf} : F \times T \longrightarrow V$,
$\mathit{CausesValue} \subseteq A \times F \times V \times T$ and
$\mathit{PossVal} \subseteq F \times V$.
To describe the general relationship between these predicates, two auxiliary predicates are defined:
$\mathit{ValueCaused} \subseteq F \times V \times T$,
$\mathit{OtherValCausedBetween} \subseteq F \times V \times T \times T$. Here,
$\mathit{ValueCaused(F, V, T)}$ means that some action happens at $T$ that gives cause for $F$ to take value $V$, and $\mathit{OtherValCausedBetween(F, V, T1, T2)}$ means that some action happens at some point in the half-open interval $[T1, T2)$ that gives cause for $F$ to take a value other than $V$. Note that ``gives cause'' is a weaker notion than the standard ``causes'': non-deterministic actions do not cause specific predictable effects. For example, rolling a die gives cause for each number to be shown, but we cannot predict which number will be shown. The FEC then consists of the following axioms:
\begin{equation*}
\text{ValueCaused} (f, v, t) \stackrel{\text{def}}{\equiv} \exists a \Big[\text{Happens}(a, t) \wedge \text{CausesValue}(a, f, v, t)\Big] \quad \text{(FEC1)}
\end{equation*}
\begin{multline*}
\text{OtherValCausedBetween}(f, v, t_1, t_2) \stackrel{\text{def}}{\equiv}\\
\exists t, v_0 \Big[\text{ValueCaused}(f, v_0, t) \wedge t_1 \leq t < t_2 \wedge v \neq v_0 \Big] \quad \text{(FEC2)}
\end{multline*}
\begin{multline*}
\left[\left(\text{ValueOf} (f, t_1)=v  \vee \text{ValueCaused}(f, v, t_1)\right) \wedge t_1 < t_2 \right.\\ 
\left.\wedge \neg \text{OtherValCausedBetween}(f, v, t_1, t_2)\right] \Rightarrow \text{ValueOf} (f, t_2)=v \quad (\text{FEC3})
\end{multline*}
\begin{multline*}
  \left[t_1 < t_2 \wedge \text{OtherValCausedBetween}(f, v, t_1, t_2) \wedge \right. \\
\quad\left. \neg \exists t(t_1 \leqslant t < t_2 \wedge \text{ValueCaused}(f, v, t))\right] \Rightarrow \text{ValueOf} (f, t_2) \neq v \quad (\text{FEC4})
\end{multline*}
\begin{equation*}
    \text{ValueOf} (f, t)=v \Rightarrow \text{PossVal}(f, v)\quad (\text{FEC5})
\end{equation*}

The notions of cause, effect, and inertia are captured in the first two FEC axioms. Axiom (FEC3) states that a fluent has a particular value at a particular time if (i) it already had that value at an earlier time or (ii) was given cause to take that value from an earlier time, and in the meantime (including that earlier time) nothing has happened that might give cause for it to take an alternative value. Conversely, (FEC4) states that fluent $f$ cannot have value $v$ at time $t_2$ if its most recent causal influences prior to $t_2$ do not include a cause for $v$. Finally, (FEC5) additionally constrains each fluent's value to be at all times among the set of values defined by $\text{PossVal}$.


The \textbf{Discrete Event Calculus} is a version of the Event Calculus that models time as a discrete quantity, with events happening at specific points in time \citep{DEC}. In the discrete Event Calculus, the duration of events is not represented and events are assumed to happen instantaneously. This makes the discrete Event Calculus well-suited for modeling systems where the duration of events is not important. 

\section{Discrete Functional Event Calculus}\label{s:DFEC}
In this paper, we introduce a variant of the discrete Event Calculus that includes non-boolean fluents: the Discrete Functional Event Calculus (DFEC). We'll start from the DEC axioms described by \citeN{DEC} and in a way similar to how FEC extends the EC, the DFEC extends the DEC to allow for non-boolean fluents. The key predicates and functions of the Discrete Functional Event Calculus are:
\begin{itemize}
    \item $\mathit{Happens(a, t)}:$ action $a$ occurs at time $t$.
    \item $\mathit{ValueOf(f, t)}$: the value of fluent $f$ at time $t$.
    \item $\mathit{PossVal(f, v)}$: value $v$ is a possible value of fluent $f$.
    \item $\mathit{CausesValue(a, f, v)}$: action $a$ causes fluent $f$ to take value $v$.
    \item $\mathit{ValueCaused(f, v, t)}$: at timepoint $t$, fluent $f$ has a cause to take value $v$
    \item $\mathit{Initiated(f, v, t)}$: fluent $f$ will have value $v$ and not be released from the commonsense law of inertia at $t+1$.
    \item $\mathit{Terminated(f, v, t)}:$ fluent $f$ will no longer have value $v$ and not be released from the commonsense law of inertia at $t+1$.
    \item $\mathit{Releases(a, f)}$: if action $a$ occurs, fluent $f$ will be released from the commonsense law of inertia.
    \item $\mathit{ReleasedAt(f, t)}$: fluent $f$ is released from the commonsense law of inertia at time $t$.
    \item $\mathit{Trajectory(f_{1}, v_{1}, t_{1}, f_{2}, v_{2}, t_{2})}$: if value $v_{1}$ of fluent $f_{1}$ is initiated by an action that occurs at timepoint $t_{1}$ , and $t_{2} > 0$, then fluent $f_{2}$ will have value $v_{2}$ at timepoint $t_{1} + t_{2}$.
    \item $\mathit{AntiTrajectory(f_{1}, v_{1}, t_{1}, f_{2}, v_{2}, t_{2})}$: if value $v_{1}$ of fluent $f_{1}$ is terminated by an action that occurs at timepoint $t_{1}$ , and $t_{2} > 0$, then fluent $f_{2}$ will have value $v_{2}$ at timepoint $t_{1} + t_{2}$.
\end{itemize}
The $\mathit{Trajectory}$ predicate, first proposed by \citeN{shanahan1990continious}, is used to capture
continuous change. If we would, for example, want to model the water level in a sink under a tap with flow rate $FR$, we could specify:
\begin{equation*}
\mathit{valueOf(Level,t_{1})}=l_{1} \Rightarrow \mathit{trajectory(Tap,Open,t_{1},Level,l_{1}+FR*t_{2},t_{2})}
\end{equation*}
The $\mathit{AntiTrajectory}$ predicate is analogous to the $\mathit{Trajectory }$ predicate, except that it is brought into play when a  value of a fluent is terminated rather than initiated by an occurring event.
\begin{figure}
\begin{equation*}
    \text{ValueCaused}(f, v, t)\stackrel{\text{def}}{\equiv}\exists a\left[\text{Happens}(a, t)\wedge \text{CausesValue}(a, f, v, t)\right]\quad \text{(DFEC1)}
\end{equation*}
\begin{equation*}
    \text{Initiated}(f, v, t) \stackrel{\text{def}}{\equiv} \text{ValueCaused}(f, v, t) \land \text{ValueOf}(f, t) = v_{1} \land v_{1} \neq v \quad \text{(DFEC2)}
\end{equation*}
\begin{equation*}
    \text{Terminated}(f, v1, t) \stackrel{\text{def}}{\equiv} \text{ValueCaused}(f, v, t) \land \text{ValueOf}(f, t) = v_{1} \land v_{1} \neq v \quad \text{(DFEC3)}
\end{equation*}
\begin{equation*}
    \text{StoppedIn}(t_{1},f,v,t_{2})\stackrel{\text{def}}{\equiv} \exists t(t_{1} < t < t_{2} \land \text{Terminated}(f,v,t)) \quad \text{(DFEC4)}
\end{equation*}
\begin{equation*}
\text{StartedIn}(t_{1},f,v,t_{2})\stackrel{\text{def}}{\equiv} \exists t (t_{1} < t < t_{2} \land \text{Initiated}(f,v,t)) \quad \text{(DFEC5)}
\end{equation*}
\begin{multline*}
\text{Initiated}(f_{1},v_{1},t_{1}) \land 0<t_{2} \land \text{Trajectory}(f_{1},v_{1},t_{1},f_{2},v_{2},t_{2}) \\ \land \neg \text{StoppedIn}(t_{1},f_{1},v_{1},t_{1}+t_{2}) \Rightarrow \text{ValueOf}(f_{2},t_{1}+t_{2})=v_{2} \quad \text{(DFEC6)}
\end{multline*}
\begin{multline*}
\text{Terminated}(f_{1},v_{1},t_{1}) \land 0<t_{2} \land \text{AntiTrajectory}(f_{1},v_{1},t_{1},f_{2},v_{2},t_{2}) \\ \land \neg \text{StartedIn}(t_{1},f_{1},v_{1},t_{1}+t_{2}) \Rightarrow \text{ValueOf}(f_{2},t_{1}+t_{2})=v_{2} \quad \text{(DFEC7)}
\end{multline*}
\begin{multline*}
    \text{ValueOf}(f,t)=v\land\neg\text{ReleasedAt}(f, t+1) \land \neg \text{Terminated}(f,v,t) \\ \Rightarrow \text{ValueOf}(f,t+1)=v \quad \text{(DFEC8)}
\end{multline*}
\begin{multline*}
    \text{ReleasedAt}(f,t) \land \neg (\text{Initiated}(f,v,t) \lor  \text{Terminated}(f,v,t)) \\ \Rightarrow \text{ReleasedAt}(f,t+1) \quad \text{(DFEC9)}
\end{multline*}
\begin{multline*}
    \neg \text{ReleasedAt}(f,t) \land \neg \exists a (\text{Happens}(a,t) \land  \text{Releases}(a,f)) \\ \Rightarrow \neg \text{ReleasedAt}(f,t+1) \quad \text{(DFEC10)}
\end{multline*}
\begin{equation*}
    \text{Initiated} (f, v, t) \Rightarrow \text{ValueOf}(f,t+1)=v \quad (\text{DFEC11})
\end{equation*}
\begin{equation*}
    \text{Terminated} (f, v, t) \Rightarrow \text{ValueOf}(f,t+1)\neq v \quad (\text{DFEC12})
\end{equation*}
\begin{equation*}
    \text{Happens}(a, t) \land \text{Releases}(a, f) \Rightarrow \text{ReleasedAt}(f, t+1) \quad (\text{DFEC13})
\end{equation*}
\begin{equation*}
    \text{Initiated}(f, v, t) \lor \text{Terminated}(f, v, t) \Rightarrow \neg \text{ReleasedAt}(f,t+1) \quad (\text{DFEC14})
\end{equation*}
\begin{equation*}
    \text{ValueOf}(f, t) = v \Rightarrow \text{PossVal}(f, v) \quad (\text{DFEC15})
\end{equation*}
\caption{Axioms of the Discrete Functional Event Calculus}
\label{fig:DFEC}
\end{figure}

The commonsense law of inertia \citep{shanahan1997solving} states that a fluent’s value persists unless the fluent is affected by an event. When a fluent is released from this law, its value can fluctuate. Fluents that are released from the commonsense law of inertia can be used to model nondeterministic effects \citep{shanahan1999event} and indirect effects \citep{shanahan1999ramification}.

Let \emph{DFEC} be the conjunction of definitions (DFEC1)-(DFEC5) and axioms (DFEC6)-(DFEC15) in Figure \ref{fig:DFEC}. Note that free variables are implicitly universally quantified and the notation $\Gamma_{1} \stackrel{\text{def}}{\equiv}\Gamma_{2}$ defines $\Gamma_{1}$ as an abbreviation for $\Gamma_{2}$. That is, $\Gamma_{1} \stackrel{\text{def}}{\equiv}\Gamma_{2}$ means that all occurrences of the expression $\Gamma_{1}$ are to be replaced with the expression $\Gamma_{2}$.

Here is a short explanation of the axioms:
\begin{itemize}
    \item \textbf{DFEC6}: If value $v_{1}$ is initiated for fluent $f_{1}$ at $t_{1}$ and this value is not stopped between $t_{1}$ and $t_{1}+t_{2}$ with  $t_{2} > 0$, then fluent $f_{2}$ has value $v_{2}$ at $t_{1}+t_{2}$.
    \item \textbf{DFEC7}: If value $v_{2}$ is terminated for fluent $f_{1}$ at $t_{1}$ and this value is not started between $t_{1}$ and $t_{1}+t_{2}$ with  $t_{2} > 0$, then fluent $f_{2}$ has value $v_{2}$ at $t_{1}+t_{2}$.
    \item \textbf{DFEC8}: If a fluent $f$ has value $v$ at timepoint $t$, the fluent is not released from the commonsense law of inertia at $t + 1$, and value $v$ of fluent $f$ is not terminated at $t$, then the fluent has the same value $v$ at $t + 1$.
    \item \textbf{DFEC9}: If a fluent is released from the commonsense law of inertia at timepoint $t$ and no value is either initiated or terminated by any action that occurs at $t$, then the fluent is released from the commonsense law of inertia at $t+1$.
    \item \textbf{DFEC10}: If a fluent is not released from the commonsense law of inertia at timepoint $t$ and the fluent is not released by any action that occurs at $t$, then the fluent is not released from the commonsense law of inertia at $t+1$.
    \item \textbf{DFEC11}: If value $v$ of fluent $f$ is initiated at timepoint $t$, then the value of fluent $f$ is $v$ at timepoint $t+1$.
    \item \textbf{DFEC12}: If value $v$ of fluent $f$ is terminated at timepoint $t$, then the value of fluent $f$ is no longer $v$ at timepoint $t+1$.
    \item \textbf{DFEC13}: If a fluent is released by some action that occurs at timepoint $t$, then the fluent is released from the commonsense law of inertia at $t + 1$.
    \item \textbf{DFEC14}: If there is value $v$ of fluent $f$ that is initiated or terminated at timepoint $t$, then the fluent is not released from the commonsense law of inertia at $t+1$.
    \item \textbf{DFEC15}: If a fluent has a value at timepoint, then this value should be a possible value of this fluent.
\end{itemize}
The DFEC can find application in diverse areas, including planning, cognitive robotics, legal reasoning and software engineering, among others. It is specifically well suited for situations where the duration of events is not important and fluents can take on several values.
Our payroll management application does not require all the features of this DFEC. In particular, in our setting, we do not need to model continuous change, with one exception, namely, that we need to keep track of the passage of time. Rather than using the general \emph{Trajectory}-predicate for this, we will find it easier to introduce a special-purpose kind of fluent, which we will call \emph{count fluents} to track for how long fluents have had their values. Consequently, besides \emph{Trajectory} and \emph{AntiTrajectory}, \emph{Releases} and \emph{ReleasedAt} predicates are deemed unnecessary. Our implementation will thus adopt a simplified version of the Discrete Functional Event Calculus, detailed in Figure 3 with the relevant definitions and axioms.
\begin{figure}
\begin{equation*}
    \text{ValueCaused}(f, v, t)\stackrel{\text{def}}{\equiv}\exists a\left[\text{Happens}(a, t)\wedge \text{CausesValue}(a, f, v, t)\right]\quad \text{(SDFEC1)}
\end{equation*}
\begin{equation*}
    \text{Initiated}(f, v, t) \stackrel{\text{def}}{\equiv} \text{ValueCaused}(f, v, t) \land \text{ValueOf}(f, t) = v_{1} \land v_{1} \neq v \quad \text{(SDFEC2)}
\end{equation*}
\begin{equation*}
    \text{Terminated}(f, v1, t) \stackrel{\text{def}}{\equiv} \text{ValueCaused}(f, v, t) \land \text{ValueOf}(f, t) = v_{1} \land v_{1} \neq v \quad \text{(SDFEC3)}
\end{equation*}
\begin{equation*}
    \text{ValueOf}(f,t)=v \land \neg \text{Terminated}(f,v,t) \Rightarrow \text{ValueOf}(f,t+1)=v \quad \text{(SDFEC4)}
\end{equation*}
\begin{equation*}
    \text{Initiated} (f, v, t) \Rightarrow \text{ValueOf}(f,t+1)=v \quad (\text{SDFEC5})
\end{equation*}
\begin{equation*}
    \text{Terminated} (f, v, t) \Rightarrow \text{ValueOf}(f,t+1)\neq v \quad (\text{SDFEC6})
\end{equation*}
\begin{equation*}
    \text{ValueOf}(f, t) = v \Rightarrow \text{PossVal}(f, v) \quad (\text{SDFEC7})
\end{equation*}
\caption{Axioms of the Simplified Discrete Functional Event Calculus}
\label{fig:SDFEC}
\end{figure}

\subsection{ASP implementation DFEC}
In the context of a limited domain, the integration of EC and DEC formulas into Answer Set Programming (ASP) programs is feasible, as demonstrated by \citeN{kim2009circumscriptive}. They showed that it is possible to embed circumscriptive theories into stable models and thus that ASP is a viable approach to computing circumscriptive theories like EC. Our implementation of the DFEC axioms in ASP, shown in \ref{appendix:DFEC}, is closely related to their implementation of the DEC axioms.  

The Discrete Functional Event Calculus (DFEC) implementation is designed to yield, at most, one solution. This deterministic outcome is rooted in the stratified nature of the DFEC. Specifically, when provided with a coherent set of fluents, actions, and effects, the DFEC implementation guarantees there is precisely one solution.
A normal logic program $\Pi$ is stratified if there exists a level mapping $f$ assigning each ground predicate symbol a natural number such that, for every rule $r$ $\in$ $\Pi$ and every predicate $P_1$ and $P_2$,
\begin{enumerate}
    \item if $P_1$ occurs in $head(r)$ and $P_2$ occurs in $body^+(r)$ then f($P_1$) $\geq$ f($P_2$);
    \item if $P_1$ occurs in $head(r)$ and $P_2$ occurs in $body^-(r)$ then f($P_1$) $>$ f($P_2$).
\end{enumerate}
To prove that the ASP implementation of the DFEC is stratified, we introduce such a level mapping. Assigning a level is based on the structure of each atom in the program. Specifically, atoms of the form $\mathit{stoppedIn(J, F, V, I)}$ or $\mathit{startedIn(J, F, V, I)}$ are assigned level $I$. Atoms of the form $\mathit{initiated(F,V,I)}$, 
$\mathit{terminated(F,V,I)}$, $\mathit{releasedAt(F,V,I)}$ and $\mathit{valueOf(F,V,I)}$, are assigned level $I+1$, while all other atoms are assigned level 0. It can be easily confirmed that this mapping is correct. As an example, consider axiom (DFEC6). In this rule, the level of the head $T1+T2+1$ is demonstrated to be greater than or equal to the level $T1$ of the positively occurring atom $\mathit{initiated(F1,V1,T1)}$. Moreover, it is strictly greater than the level of the negatively occurring atom 
\\$\mathit{stoppedIn(T1,F1,V1,T1+T2)}$. Similar verification can be conducted for the remaining axioms, thereby establishing the correctness of the level mapping. The level mapping of the $DFEC$ implementation is provided in \ref{appendix:Mapping}.

\section{Proposed Approach}\label{proposedapproach}
We will now outline our general approach to tackling the three challenges described in the introduction.

To tackle the first challenge, we base ourselves on a generic implementation of the Discrete Functional Event Calculus defined in Section \ref{s:DFEC}. The HR experts can then represent the rules for a specific company by defining a concrete set of actions and fluents. Facilitating a user-friendly approach to this task, we leverage Decision Model and Notation (DMN) decision tables to define actions, fluents, and consequently, the payroll rules. This approach empowers HR experts to understand the underlying logic of the payroll system and enables them to take on most of the work of maintaining it. 
However, the DMN notation has no notion of time and thus temporal properties cannot be expressed in standard DMN. To provide the necessary expressive power, we extend this notation using simple temporal logic and interval logic, which we will describe below. 

To cope with the second challenge, we define the semantics of the language that is provided to the HR experts by means of a declarative Answer Set Program. Known for its flexibility and elaboration tolerance, ASP allows us to implement this language in such a way that new language features can easily be added when necessary, with minimal running risk of introducing bugs in the existing models. Once the HR experts articulate the payroll rules within DMN tables, the system can then automatically transform these tables into an ASP program.

One downside of using the expressive declarative ASP formalism is that the third challenge, i.e., that of computational efficiency, may be hard to meet. Indeed, as we show below, the usual ground-and-solve approach of ASP solvers falls far short of the required performance. To address this issue, we make use of the multi-shot solving capabilities of the ASP solver \texttt{clingo}. Thus, upon the translation of the tables into ASP, the resulting program is handed over to \texttt{clingo} for efficient processing and execution of the payroll operations.

We now define the temporal and interval logic to extend the DMN notation with the possibility to express temporal properties.

\subsection{Temporal Logic}\label{logictimepoints}
Some of the domain knowledge that the consultants need to express concerns temporal properties. For instance, overtime starts after an employee has been working for 8 hours.
We represent such knowledge using the following simple linear time logic:
\begin{itemize}
\item An atomic temporal formula is of the form $f=v$, with $f$ a fluent and $v$ a value from the domain of $f$; 
\item If $\phi$ and $\psi$ are temporal formulas, then so are $\phi \land \psi$ and $\lnot\phi$; 
\item If $\phi$ is a temporal formula and $n\in \mathbb{N}$, then $[\geq n]\phi$ is a temporal formula, which intuitively represents that $\phi$ has been true for at least the $n$ most recent time points, including the present one.
\end{itemize}
Given a time line $T = (I_0,I_1, \ldots)$ and a time point $i\geq 0$, we define that a temporal formula $\phi$ holds in $(T,i)$, denoted as $(T,i)\models \phi$, as follows:
\begin{itemize}
\item For an atomic temporal formula, $(T,i) \models f=v$ if $f^{I_i}=v$ i.e., the value of fluent $f$ in the state of the world $I_i$ at time point $i$ is equal to $v$;
\item For a conjunction $\phi\land\psi$, $(T,i) \models \phi\land \psi$ if $(T,i) \models \phi$  and $(T,i) \models \psi$;
\item For a negation $\lnot\phi$, $(T,i) \models \lnot\phi$ if $(T,i) \not \models \phi$ ;
\item For $[\geq n] \phi$, $(T,i)\models [\geq n]\phi$ if $i \geq n$ and for all $i-n < j \leq i$ , $(T,j) \models \phi$.
\end{itemize}
We also introduce an abbreviation $[=n]\phi$ for the formula $[\geq n]\phi \land \lnot [\geq n+1] \phi$. As we will show in Section \ref{conditions}, this temporal logic is easily defined in ASP.

\subsection{Interval Logic}\label{logicintervals}
The wages of employees are not defined in terms of individual timepoints but in terms of \textbf{intervals}. The consultants can define which fluents are considered \emph{relevant} for the wages of employees.
Within an interval, all the relevant fluents keep their value, i.e., the boundaries of the interval are timepoints at which the value of a relevant fluent changes. We consider half-open intervals [i,j) where $i$ is a timepoint at which the value of a relevant fluent changes and $j$ is the next such timepoint.
An \emph{interval property} describes a characteristic of an interval. It is either a relevant fluent $f$, or an aggregate function like $\mathit{length}$. We denote the value of an interval property $p$ in an interval $[i,j)$ as $p^{[i,j)}$. Given a timeline $T = (I_0,I_1, \ldots)$, we define this value in the following way:
\begin{itemize}
    \item For a relevant fluent $f$, $f^{[i,j)} = f^{I_i}$, that is, the value of fluent $f$ at the start of the interval. Because $f$ is a relevant fluent, this is also the value that $f$ has in all following $I_k$ for $k \in [i,j)$.
    \item For the aggregate function $\mathit{length}$, $\mathit{length}^{[i,j)} = j - i$.
\end{itemize}

In addition to interval properties, we also define \textbf{interval terms}. An interval term is an expression that refers to a specific interval. The atomic interval term $\mathit{this}$ refers to the current interval. For an interval term $t$, $next(t)$ refers to the next interval, and $prev(t)$ to the previous one. Given a sequence of intervals $S = (\mathcal{I}_0, \mathcal{I}_1,\ldots)$, we define $\mathit{this}^{(S,i)} = \mathcal{I}_i$ and for every interval term $t$ with $t^{(S,i)} = \mathcal{I}_j$, we define $next(t)^{(S,i)} = \mathcal{I}_{j+1}$ and  $prev(t)^{(S,i)} = \mathcal{I}_{j-1}$.

An \textbf{interval value} is then of the form $[t]p$ and refers to the value of the interval property $p$ for the interval that is indicated by the interval term $t$. We define this value as $[t]p = p^{t^{(S,i)}}$. 
We define an \textbf{atomic interval formula} as $v \theta w$ where $v$ and $w$ are two interval values and $\theta$ is a comparison operator. Finally, we combine these atomic interval formulas with the standard boolean operators, as usual.

\section{Single-shot implementation}\label{ssimplementation}

First, we introduce an implementation designed for utilization with a conventional single-shot solver. Afterwards, we show how a more efficient implementation can be developed that uses the multi-shot solving capabilities of \emph{clingo}.

\subsection{Discrete Functional Event Calculus}\label{impl_DFEC}
The core of our generic model is an implementation of the simplified version of the Discrete Functional Event Calculus described in Section \ref{s:DFEC}. The timeline considered in our model consists of a number of discrete timepoints $0..n$, represented by the \emph{time(T)} predicate. Each timepoint corresponds to a certain duration in wall-clock time. For the moment, we consider timepoints that are 10 minutes long.
The \emph{ts(Days, Hours, Minutes, Stamp)} predicate links a duration expressed in real-world $\mathit{Days}$, $\mathit{Hours}$ and $\mathit{Minutes}$ to the corresponding timepoint $\mathit{Stamp}$ (lines 2-5).

\hspace{-1.1em}
\begin{minipage}{\linewidth}
\begin{lstlisting}[]
day(0..1).  hour(0..23).    minute(0..59).
ts(Days,Hours,Minutes,Stamp):-day(Days),hour(Hours),minute(Minutes),
    Stamp = (Days*24+Hours)*6+Minutes/10.
time(T):-day(Days),hour(Hours),minute(Minutes),
    ts(Days,Hours,Minutes,T).
maxTime(M):-M = #max{T: time(T)}.
\end{lstlisting}
\end{minipage}
The situation of the employee at a certain time point is described by a complete set of \textbf{fluents}. Fluents have a domain of possible values. 
An example is the boolean fluent \emph{present}, which indicates whether the employee is currently clocked in. 
At each point in time, a fluent has a certain value, as represented by the predicate \emph{value(F,V,T)}. At timepoint 0, the fluent's value is specified by the \emph{initially(F,V)} predicate. 
The \emph{cause(F,V,T)} predicate represents that at timepoint $T$ there is a cause for the value of $F$ to change to $V$, which leads to the current value of $F$ being terminated at $T+1$ and the value $V$ being initiated (lines 11-14).

\begin{lstlisting}[escapechar=@]
value(F,V,0):-fluent(F),initially(F,V),domain(F,V).
value(F,V,T+1):-fluent(F),time(T+1),value(F,V,T),
    not terminated(F,V,T). @\textit{(SDFEC4)}@
value(F,V,T+1):- fluent(F),time(T+1), initiated(F,V,T). @\textit{(SDFEC5)}@
initiated(F,V,T):- fluent(F),cause(F,V,T),value(F,V2,T), 
    V2 != V. @\textit{(SDFEC2)}@
terminated(F,V,T):- fluent(F),value(F,V,T),cause(F,V2,T), 
    V2 != V. @\textit{(SDFEC3)}@
:- terminated(F,V,T),value(F,V,T+1). @\textit{(SDFEC6)}@
:- value(F,V,T),not PossVal(F,V). @\textit{(SDFEC7)}@
\end{lstlisting}
The only reason for a fluent to change is because an \textbf{action} changes it. In general, the effect of an action may depend on the state of the world at the time the action occurs. However, for the moment, we will consider only actions that have the fixed effect of causing a \emph{Fluent} to take on a specific \emph{Value}, as represented by the \emph{causes(Action,Fluent,Value)} predicate on line 21. 
For now, we define 2 types of actions. A \textbf{user action} is an action that is performed by the user in an intentional way (i.e., the employee). A \textbf{wall-time action} happens at a specific wall-clock time.
Both wall-time actions and user actions thus happen at a given absolute time, represented by the \emph{userDoes(A,T)} and \emph{actionTime(A,T)} predicates. 

\begin{lstlisting}[escapechar=@]
action(A):-userAction(A).
action(A):-walltimeAction(A).
happens(A,T):-userAction(A),userDoes(A,T),time(T).
happens(A,T):-walltimeAction(A),actionTime(A, T),time(T).
cause(F,V,T):-action(A),happens(A,T),causes(A,F,V,T), 
    time(T). @\textit{(SDFEC1)}@
\end{lstlisting}
With these concepts, we now represent the case-specific knowledge in an easy-to-read tabular representation. The case-specific knowledge for our running example is shown in Fig. \ref{fig:eventcalculustable}. Each table enumerates a single predicate, e.g., the first table corresponds to the $\mathit{initially(F,V)}$ predicate, etc.


\begin{figure}\scriptsize
    \centering
    \dmntable{Inertial Fluent}{}{Inertial Fluents}{Initial value}
             {present, false,
              timeOfDay, night,
              dayStarted, false,
              restBreakPossible, false,
              shifStarted, false,
              cumul,false,
              break,no}
    \dmntable{Actions effects}{}{Action,Affected Fluent}{Action causes Fluent to}
            {clockIn, present, true,
             clockout, present, false,
             nightfall, timeOfDay, night,
             morning, timeOfDay, day,
             dayStarts, dayStarted, true,
             dayEnds, dayStarted, false
            }
    \dmntable{Unconditional action effects}{}{Action,Fluent}{Action causes Fluent to}
            {...,...,...,
             clockIn, present, true,
             makeRestBreakPossible,restBreakPossible,true,
             ...,...,...
            }
    \dmntable{Walltime Actions}{}{Action}{Wall-time}
            {
             nightfall,22h00,
             morning,7h00,
             dayStarts,14h00,
             dayEnds,23h00 
            }
    \dmntable{User Actions}{}{Action}{User does Action at}
            {
             clockIn,13h45,
             clockOut,16h00,
             clockIn,16h30,
             clockOut,23h30 
            }
    \caption{Case-specific representation of a basic scenario}
    \label{fig:eventcalculustable}
    
\end{figure}

\subsection{Conditional Effects}\label{conditions}
The effect of an action may depend on the current state of the world.
We describe such a conditional effect with a \emph{ccauses(Action, Fluent, Value, Condition)} predicate.
\begin{lstlisting}[]
causes(A,F,V,T):-ccauses(A,F,V,Cond),holds(Cond,T).
\end{lstlisting}
Conditions are formatted in the temporal logic of Section \ref{logictimepoints}. In the tables, we use a slightly more user-friendly syntax, writing, e.g., 
\begin{equation*}
f1=v1 \wedge [=n] f2=v2 \quad \text{as} \quad f1=v1\ and\ f2 = v2 \ since\ n.
\end{equation*}
In our ASP implementation, we represent each such condition as a set of facts. For instance, the above condition $\mathit{cond1}$ is represented as follows:

\begin{lstlisting}[]
and(cond1,2).   sub(cond1,0,v(f1,v1)).  sub(cond1,1,since(f2,v2,n)).
\end{lstlisting}
Here the $\mathit{and(C,N)}$ predicate denotes that condition $C$ is a conjunction of $N$ sub-conditions, each represented by a $\mathit{sub(C,I,SC)}$ fact, with $0\leq I<N$.
The $\mathit{holds(Cond,T)}$ predicate specifies whether a condition $\mathit{Cond}$ holds at a certain timepoint $T$. 
For $\mathit{Cond}=v(F,V)$, we just check whether Fluent $F$ has Value $V$ (line 25). $\mathit{Cond=since(F,V,D)}$, representing whether a fluent $F$ has had a value $V$ for $D$ timepoints, is defined by two rules. The first rule states that the initiation of $V$ for $F$ at $T-1$ marks timepoint $T$ as the start of a period in which $F$ has value $V$ (line 26). The recursive rule on line 27 extends such a period with one timepoint if $F$ still has value $V$ at $T$. 
Finally, if $\mathit{Cond}$ is a conjunction, we check whether the number of conjuncts that hold at $T$ matches its total number of conjuncts $N$ (line 28).

\hspace{-1.1em}
\begin{minipage}{\linewidth}
\begin{lstlisting}[]
holds(v(F,V),T):-value(F,V,T).
holds(since(F,V,0),T):-time(T),initiated(F,V,T-1).
holds(since(F,V,D+1),T):-time(T),value(F,V,T),
    holds(since(F,V,D),T-1).
holds(C,T):-and(C,N),time(T),#count{I:sub(C,I,Sub),holds(Sub,T)}=N.
\end{lstlisting}
\end{minipage}
We can now represent scenarios where these conditional effects come into play in the tabular representation. In the first table of Figure \ref{fig:conditionaleffecttable}, we specify that a shift starts if an employee clocks in when the workday has already started or when the day starts and the employee is already clocked in.
\begin{figure}\scriptsize
    \centering
    \dmntable{Conditional effects}{}{Action, Condition, Fluent}{\makecell{Action changes Fluent to \\ if Condition holds}}
            {clockIn, dayStarted=true,
              shiftStarted, true,
              dayStarts, present=true,
              shiftstarted, true
            }
    \dmntable{Triggered Action}{}{Action}{Condition that triggers Action}
            {makeRestBreakPossible,shiftStarted=true since 1h00}
    \caption{Case-specific representation of conditional effects and triggered actions}
    \label{fig:conditionaleffecttable}
\end{figure}
\subsection{Automatically triggered actions}
An action can also be automatically triggered if a certain temporal condition is fulfilled. To specify that an employee's break is paid as an (official) rest break after their shift has lasted for one hour, we specify an action \emph{makeRestBreakPossible}, which causes fluent \emph{restbreakPossible} to be \emph{true}. 
Such actions are not performed by the user but happen automatically after a certain fluent $F$ has had a certain value $V$ for a specific duration $D$, as represented by an \emph{after(F,V,D,A)} predicate (line 31). Figure \ref{fig:conditionaleffecttable} represents triggered action \emph{makeRestBreakPossible} in our tabular form. 
\begin{lstlisting}[]
action(A):-triggeredAction(A).
happens(A,T):-holds(since(F,V,Dur),T),after(F,V,Dur,A),time(T).
\end{lstlisting}
\subsection{Defined fluents}
To avoid repeated use of complex fluent formulas, we introduce \textbf{defined fluents} as defined by \citeN{denecker2007inductive} and used in action languages like $\mathcal{AL}_d$ \citep{gelfond2009yet}. 
The value of a defined fluent is completely determined by the values of the other fluents. Therefore, they provide no additional information about the state of the world; they simply make it easier to track its properties. The value of a defined fluent is defined by a set of $\mathit{rule(F,V,C)}$ facts: if condition $C$ holds, the defined fluent $F$ has the value $V$ (line 32). To cover the case that no rules are applicable, each defined fluent must have a default value (line 33). We do not allow fluents to be defined by negated predicates, therefore are no cycles in our solution. As stated by \citeN{propertiesALd}, acyclic behaviour arises when there are no paths from defined fluents to their negations. Given that acyclic behaviour is a sufficient condition for well-foundedness, we can deduce that our defined fluents will consistently have, at most, one value.

\hspace{-1.1em}
\begin{minipage}{\linewidth}
\begin{lstlisting}[]
value(F,V,T):-defined(F),time(T),rule(F,V,C),holds(C,T).
value(F,V,T):-defined(F),time(T),default(F,V),not appliedRule(F,T).
appliedRule(F,T):-rule(F,V,C),holds(C,T).
\end{lstlisting}
\end{minipage}
For example, to indicate whether an employee is working, we introduce the defined fluent $\mathit{atWork=true}$ if and only if the inertial fluents \emph{present} and  \emph{shiftStarted} are \emph{true}. Figure \ref{fig:definedtable} represents this in our tabular representation.

\begin{figure}\scriptsize
    \centering
    \dmntable{Defined Fluents}{}{Defined Fluent}{Default Value}
            {atWork,false
            }
    \dmntable{Count Fluent}{}{Count Fluent}{Count Rule}
            {workedHours,atWork=true
            }
    \dmntable{Defined Fluent}{}{Defined Fluent,Rule}{Value}
            {
             atWork,present=true and shiftStarted=true,true
            }
    \centering
    \caption{Case-specific representation of defined and count fluents}
    \label{fig:definedtable}
\end{figure}
\subsection{Count fluents}
An employee receives a bonus for any overtime done after they have already worked eight hours. This can be modelled with a third type of fluents, the \textbf{count fluents}. These fluents keep track of the past states of the world by counting the foregoing timepoints in which a certain condition holds, upon a certain timepoint.
Each count fluent is defined by a rule of the form $\mathit{countRule(CF,Cond)}$, which specifies that the fluent $\mathit{CF}$ counts the timepoints at which the condition $\mathit{Cond}$ holds (lines 36-39).

\begin{lstlisting}[]
value(CF,0,0):-countFluent(CF).
value(CF,S+1,T):-countFluent(CF),value(CF,S,T-1),countRule(CF,Cond),
    holds(Cond,T).
value(CF,S,T):-countFluent(CF),value(CF,S,T-1),countRule(CF,Cond),
    not holds(Cond,T).
\end{lstlisting}
In Fig. \ref{fig:definedtable}, we define the count fluent \emph{workedHours}, which, at each timepoint, keeps track of how long the fluent \emph{atWork} has been $\mathit{true}$ up to that point.

Count fluents can also trigger actions. The predicate $\mathit{when(CF,V,A)}$ states that if count fluent $\mathit{CF}$ reaches a certain value $\mathit{V}$, an action $A$ is triggered.
To make sure that an action is not triggered multiple times if the desired value is maintained for multiple timepoints, we state that the value of the count fluent in the previous timepoint should be smaller than the desired value (line 40).
\begin{lstlisting}[]
happens(A,T):-when(CF,V,A),value(CF,V,T),value(CF,V1,T-1),V>V1.
\end{lstlisting}
For example, the overtime bonus is applied by triggering an action \emph{cumulPremiumAction} after eight hours of work. Figure \ref{fig:countTriggeredTable}, adds this triggered action and its effect to the existing tables for triggered actions and effects (Figures \ref{fig:conditionaleffecttable} and \ref{fig:eventcalculustable}). Note that conditions specifying that a fluent $F$ has had a value $V$ for a duration $D$ are translated in ASP as an $after(F,V,D,A)$ predicate and those that specify value $V$ for a count fluent $CF$ are translated as a $when(CF,V,A)$ predicate.
\begin{figure}\scriptsize
    \centering
    \dmntable{Triggered Action}{}{Action}{Condition that triggers Action}
            {..., ...,
            cumulPremium,workedHours=8h00
            }
   \dmntable{Actions effects}{}{Action,Fluent}{Action causes Fluent}
            {...,...,...,
             cumulPremium,cumul,true
            }
    \caption{Extended case-specific representation of triggered actions and effects}
    \label{fig:countTriggeredTable}
\end{figure}

\subsection{Reasoning about intervals}\label{intervals}
Up to now, we have only reasoned about individual timepoints. To calculate the total wages of employees we consider intervals. As described in Section \ref{proposedapproach}, within intervals all of the ``relevant" fluents keep their value. By restricting attention to only the relevant fluents, we avoid creating too many small intervals. A relevant fluent is annotated by the \emph{relevant(Fluent)} predicate.  We denote the relevant fluents by introducing a new column in the inertial and defined fluent table (Figures  \ref{fig:eventcalculustable} and \ref{fig:definedtable}), as can be seen in Figure \ref{fig:relevanttable}.
\begin{figure}\scriptsize
    \centering
    \dmntable{Inertial Fluents}{}{Inertial Fluents}{Default Value, Relevant}
             {
              timeOfDay, night, true,
              cumul, false, true,
              ..., ..., ...
              }
    \dmntable{Defined Fluent}{}{Defined Fluent}{Default Value, Relevant}
            {
             atWork,false,true
            }
    \caption{Case-specific representation of relevant fluents}
    \label{fig:relevanttable}
    \vspace{-1em}
\end{figure}
The boundaries of intervals are those timepoints at which the value of at least one relevant fluent changes (lines 41-42). 

\hspace{-1.1em}
\begin{minipage}{\linewidth}
\begin{lstlisting}[]
boundary(T):-relevant(F),changes(F,T).
changes(F,T):-time(T),time(T-1),value(F,V1,T-1),value(F,V2,T),V1!=V2.
\end{lstlisting}
\end{minipage}
To define the intervals, we assign an $id$ to each boundary. The intervals are of the form $[B(i),B(i+1))$, where $B(i)$ denotes the boundary with id $i$. The \emph{stretchesTo(Id,T)} predicate denotes that interval $Id$ includes timepoint $T$. We model this with two rules. The first one states that if the interval includes the previous timepoint $T-1$ and timepoint $T$ is not a boundary then the interval includes $T$ as well (line 47). Boundaries themselves are included in the time interval that they start (line 48). Timepoint $T$ is thus a boundary of time interval $I+1$ if timepoint $T-1$ is included in the previous interval $I$ and $T$ is a boundary (line 46). A boundary thus denotes the start of an interval (\emph{intervalFrom(Id,From)} predicate) and the end of the previous interval (\emph{intervalTo(Id,To)} predicate) (lines 44-45).
\begin{lstlisting}[]
boundary(0,0).
intervalFrom(Id,From):-boundary(Id,From).
intervalTo(Id,To):-id(Id),boundary(Id+1,To).
boundary(I+1,T):-id(I+1),id(I),boundary(T),stretchesTo(I,T-1).
stretchesTo(I,T):-id(I),stretchesTo(I,T-1),not boundary(T). 
stretchesTo(I,T):-boundary(I,T).
\end{lstlisting}
In principle, there could be as many intervals as there are timepoints. Typically, there will be far fewer. To speed up the program, we assume an upper bound of at most 20 intervals. We require that the final interval id be not used. If this ever happens, the upper bound should be increased.
Next, as described in the interval logic of Section \ref{logicintervals}, we define interval terms, such as \emph{prev(this)} in lines 55 to 57.

\hspace{-1.1em}
\begin{minipage}{\linewidth}
\begin{lstlisting}[]
id(0..20).
enoughIds:-freeId(I).
freeId(I):-id(I), maxId(M), I > M.
maxId(M):-id(M), M = #max{I: usedId(I)}.
usedId(I):-id(I), boundary(I,T).
:-not enoughIds.
refersTo(this,Id,Id):-id(Id).
refersTo(prev(I),J-1,Id):-id(Id), id(J), refersTo(I,J,Id).
refersTo(next(I),J+1,Id):-id(Id), id(J), refersTo(I,J,Id).
\end{lstlisting}
\end{minipage}
To clearly track the characteristics of the interval and avoid repeated use of complex interval formulas, we introduce \textbf{defined interval properties}. The value of a defined property is specified by a set of \emph{intRule(P,V,C)} facts. If the condition $C$ holds in the current interval, the defined property $P$ has the value $V$ (line 72). These conditions are formatted in the interval logic of Section \ref{logicintervals}. 
The \emph{intHolds(Cond,Id)} predicate specifies when a condition for an interval holds.
To check whether a condition holds, we define the value of the properties described in the interval logic of Section \ref{logicintervals}.
A \emph{valueOfProp(P,V,Id)} predicate denotes that an interval property $P$ has value $V$ in interval $Id$. If $P$ is a relevant fluent, this is simply the case if this fluent has that value (line 58). If $P = length$, its value is the length of interval $Id$ (lines 60-62).
The value of an interval atom $\mathit{[IntTerm]P}$, which we denote in ASP as $\mathit{at(IntTerm,P)}$, is the value of a property $P$ in the interval that $\mathit{IntTerm}$ refers to. We define when two such atom values are equal (line 65), when one atom value is smaller than another (line 69), and when an atom value has a certain value from its domain (line 67). 
Finally, a conjunction $C$ holds in an interval, if all $N$ of its conjuncts, specified by the \emph{iand(C,N)} predicate, hold (lines 70-71).
   


\begin{lstlisting}
valueOfProp(F,V,Id):-relevant(F),value(F,V,T),id(Id),
    intervalFrom(Id,T).
valueOfProp(length,Value,Id):-id(Id), length(Id,Value).
length(Id,L):-intervalFrom(Id,From), intervalTo(Id, To),
    L = To - From.
valueOfAtom(at(IntTerm,Prop),Val,Id):-refersTo(IntTerm,Int,Id),
    valueOfProp(Prop,Val,Int).
intHolds(equals(A1,A2),Id):-valueOfAtom(A1,V,Id), 
    valueOfAtom(A2,V,Id).
intHolds(hasValue(Atom,V),Id):-valueOfAtom(Atom,V,Id).
intHolds(less(A1,A2),Id):-valueOfAtom(A1,V1,Id),
    valueOfAtom(A2,V2,Id), V1 < V2.
intHolds(C,Id):-iand(C,N),id(Id),
    #count{I : isub(C,I,S),intHolds(S,Id)} = N.
definedProp(Prop,V,Id):-id(Id),intRule(Prop,V,C),intHolds(C,Id).
\end{lstlisting}
In Figure \ref{fig:intervalfluenttable}, to specify the wages in a certain interval, we introduce some defined properties. The \emph{normalwage} property is 20 euros if an employee is working, and 0 if they are not. They receive a \emph{nightpremium} of 20\% for any interval in the night or in the day if the interval precedes an interval in the night, provided that the night interval is larger. Finally, the employee receives an \emph{cumulPremium} of 25\%. 
Note that compared to the interval logic of \ref{logicintervals} we omit the interval term $\mathit{this}$ in our tables to improve readability. 
\begin{figure}\scriptsize
    \centering
    \dmntable{Defined properties}{}{Defined property}{Default Value}
             {normalwage,0,
             nightpremium,0,
             cumulpremium,0,
             totalWage,\makecell{= normalwage $\times$ \\ (1 + nigtpremium\\+ cumulpremium)}}
    \dmntable{Defined property rules}{}{Defined property,Rule}{Value}
            {
             normalwage,atWork=true,20,
             nightPremium,\makecell{timeOfDay=day \\ and [next]timeOfDay=night \\ and length $<$ [next]length},0.20,
             nightPremium,timeOfDay=Night,0.20,
             cumulPremium,cumul=true,0.25
            }
    \caption{Case-specific representation of relevant fluents and output}
    \label{fig:intervalfluenttable}
\end{figure}

\subsection{Calculating the total wages}
    %
    %
    %
In general, our goal is to compute the wages as a sum $\sum_{interval\ i} totalWage(i)$ with the \emph{totalWage} property determining the hourly wage in each interval. 
We translate this to an ASP sum as follows:

\vspace{0.3em}
\hspace{-1.1em}
\begin{minipage}{\linewidth}
\begin{lstlisting}[]
totalWage(S):-S = #sum{T : definedProp(totalWage,W,I),length(I,L),
    T = W * L/60}.
\end{lstlisting}
\end{minipage}

In total, an HR consultant only needs 10  tables to specify the specific rules. A concrete scenario can be represented as the single-user action table. 

\section{Multi-shot implementation}\label{msimplementation}
In the previous section, we presented a model for use by a standard single-shot ASP solving. This model may produce large groundings, which form a bottleneck for realistic instances. Indeed, to handle a scenario of two days with an accuracy of one minute, for instance, we need $60 \times 48$ timepoints. Because the grounding size is quadratic in the number of timepoints, this is problematic. In this section, we show how we can use \emph{multi-shot solving} to drastically reduce the grounding size, by restricting attention to only those timepoints at which the state of the world changes. We refer to these timepoints as \textbf{changepoints}.
Our multi-shot model is purely an optimised version of the single-shot model: functionally, it is still the same, and it still allows the HR consultant to represent his knowledge in the same user-friendly way. It consists of 3 parts: a static part, a dynamic part, and an interval part.
\subsection{Static Code}
The static part of the code contains the non-temporal information, which consists of rules that contain no predicates with a timepoint argument. For example, in the last code listing of the Functional Event Calculus paragraph in Section \ref{impl_DFEC}, the first two lines, which state that \emph{user actions} and \emph{walltime actions} are two kinds of actions, are included in the static part, while the last two rules, which define \emph{at which timepoints} such actions happen, does not.
We collect the static code in a subprogram $\#$\emph{static()}, that takes no parameters.

\subsection{Dynamic Code}
The dynamic program defines the current state of the world in terms of the previous state. It therefore takes two changepoints as parameters. The program first asserts that the current changepoint $cp$ is a new timepoint, which follows the previous changepoint $pp$.
In addition to this, the dynamic program also contains all rules that include predicates that take a timepoint as an argument. These rules typically define the value of some dynamic predicate at $T+1$ in terms of the values of dynamic predicates at timepoint $T$. Such rules now undergo a minor syntactic change, where we replace all such terms $T+1$ by the parameter $cp$ of the program and the terms $T$ by its parameter $pp$. In effect, this change is what allows us to ``skip ahead'' to the next changepoint, instead of having to go through each timepoint individually. This necessitates a number of other small changes in the code, documented  online\footnote{\url{https://gitlab.com/EAVISE/bca/asp-for-flexible-payroll-management/blob/main/General_MS_Implementation.lp}}, the specifics of which we will not describe in detail.

\subsection{Multi-shot Solving Algorithm}
Algorithm \ref{solve} shows how the static and dynamic programs can be used to implement the desired behavior, employing clingo's multi-shot solving. This algorithm uses the following notations: For a program $P(x_1, \ldots x_n)$ with parameters $x_i$ and constants $c_1,\ldots,c_n$, we denote by AnswerSet$(P(c_{1},..,c_{n}))$ the unique answer set of $P(c_1,\ldots,c_n)$. For a predicate $p$, we denote by $p^X$ the set of all tuples $\vec{c}$ such that $p(\vec{c}) \in X$.

At line 1, the static program $\mathit{P_{static}}$, is solved. Its answer set provides the upper-bound \emph{max} of the timeline and the list \emph{upfrontPoints} of all changepoints that are already known up-front, i.e., all the time points at which wall-time actions or user actions happen. We also include the greatest time point \emph{max} in this list, to ensure termination of the \emph{while}-loop (line 6). In each iteration, this loop instantiates the dynamic program $\mathit{P_{dynamic}}$ for the next changepoint. The \emph{if}-test (line 8) distinguishes two cases: either the next changepoint comes from the list \emph{upfrontPoints}, or else it corresponds to a \emph{triggered} action. The next timepoints at which such a triggered action occurs are not known up-front but are computed in each iteration of this loop by the function \emph{searchNext}.

The \emph{searchNext} algorithm (Algorithm \ref{search}) considers both actions that are triggered by a fluent maintaining its value for a certain time, represented by the \emph{after}-predicate (line 3), and those triggered by a count fluent reaching a certain value, represented by the \emph{when}-predicate (line 8). The result of the algorithm is the smallest timepoint \emph{next} $>$ \emph{current} at which such an action happens (or $\infty$ if no such timepoint exists). 
Once the main \emph{while}-loop of Algorithm \ref{solve} ends, the program $\mathit{P_{dynamic}}$ has been grounded for all changepoints. The predicate $\emph{boundary}$ now identifies all the timepoints that delineate an interval. The \emph{for}-loop in line 16 then introduces an identifier $i$ for each such interval $[b_i, b_{i+1})$. Finally, these intervals are then passed to the program $\#$\emph{intervals}, which gathers all of the rules concerning intervals, unchanged from our single-shot implementation.
\begin{algorithm}
\caption{Solving algorithm}\label{solve}
\begin{algorithmic}[1]
\State $X$ = AnswerSet($P_{static}())$
\State $\mathit{max} =$ the unique $t$ such that $(t) \in \mathit{maxTime^X}$
\State $\mathit{current} = 0$
\State $\mathit{upfrontPoints}=$ sorted list of all $t$ where $t=\mathit{max}$ or $\exists a: (a,t) \in \mathit{happens^X}$
\State $i = 0$ \Comment{$i$ is the counter for action points list}
\While{$\mathit{current} < \mathit{max}$}
\State $\mathit{triggered} =$ searchNext($\mathit{current}$,$X$)
\If{$\mathit{triggered} < \mathit{upfrontPoints[i]}$}
\State $\mathit{next} = \mathit{triggered}$
\Else
\State $\mathit{next} = \mathit{upfrontPoints[i]}$
\State $i$ += 1
\EndIf
\State $X$ = AnswerSet($P_{dynamic}(current,next))\cup X)$
\State $\mathit{current} = \mathit{next}$
\EndWhile
\State $(b_1, \ldots ,b_n) =$ sorted list of all $t$ such that $(t) \in \mathit{boundary}^X$
\For{$i$ \textbf{in} $1, \ldots, n-1$}
 \State$X = X \cup \{\mathit{intervalFrom}(i,b_i),\mathit{intervalTo}(i,b_{i+1})\}$
\EndFor
\State $X$ = AnswerSet($P_{\mathit{intervals}}()\cup X$)
\end{algorithmic}
\end{algorithm}

\begin{algorithm}
\caption{searchNext algorithm}\label{search}
\begin{algorithmic}[1]
\Procedure{searchNext}{$\mathit{current}$,$X$}
\State $next=\infty$
\For{$(f,v,d,a)$ \textbf{in} $\mathit{after}^X$}
\If{$\exists t'$ = the unique $t''$ such that $(f,v,t'',\mathit{current}) \in \mathit{sameSince}^X$}
\State $\mathit{possible} = t'+d$
\If{$\mathit{current}<\mathit{possible}<\mathit{next}$} 
\State $\mathit{next}=\mathit{possible}$
\EndIf
\EndIf
\EndFor
\For{$(cf,cv,a)$ \textbf{in} $\mathit{when}^X$}
\State $\mathit{cond}$ = the unique $c$ such that $(cf,c) \in countRule^X$
\setlength\floatsep{1.25\baselineskip plus 3pt minus 2pt}
\If{$\mathit{(cond,current)} \in holds^X$}
\State $\mathit{cv'}$ = the unique $\mathit{cv''}$ such that $(\mathit{cf,cv'',current}) \in \mathit{value}^X$
\State $\mathit{possible = current+(cv-cv')}$
\If{$\mathit{current < possible < next}$}
\State $\mathit{next = possible}$
\EndIf
\EndIf
\EndFor
\State \textbf{return} $\textbf{next}$
\EndProcedure
\end{algorithmic}
\end{algorithm}

In our solution, we organized our logic program into three distinct modules: static, dynamic, and interval. Importantly, we ensured the independence of these modules by strictly confining the definition of predicates within each module. Even for the dynamic module undergoing iterative grounding and solving with the current and previous changepoint, this principle holds. The introduction of the '\emph{current} $<$ \emph{possible}' constraint in the \emph{searchNext} algorithm guarantees a logical sequence, where each changepoint is always greater than the previous one. Since there are no head predicates inferred for both the previous and current changepoint within the same iteration, this design ensures that no ground literals are inferred across different iterations. Our approach aligns with the principles outlined in the Module Theorem, as described in Section~\ref{Preliminaries}.

\section{Experimental results and discussion}\label{benchmark}

As discussed before, the grounding size of the single-shot implementation is quadratic in the number of timepoints. Consequently, the number of timepoints has a large effect on the computational performance of this approach. 
Figure \ref{results} shows how the duration of a single timepoint affects the computation time for a two-day scenario, the minimum for a realistic scenario. The company we collaborated with would like a single scenario to be handled in under a second of computation time. At the same time, a granularity in which a single time point is more than five minutes in length is unacceptable for them.  Figure \ref{results} therefore clearly shows that the single-shot implementation is not feasible.

The multi-shot approach drastically reduces the impact of the total number of timepoints on the computation time. Indeed, the main parameter is now the number of \emph{changepoints}, which means that the run-time is mainly determined by the scenario that needs to be handled, rather than by the granularity of the timepoints. Figure \ref{results} shows that if we change the granularity and the number of timepoints increases, the computation time of our multi-shot approach barely increases. Figure \ref{fig:changepointresults} shows how the computation time of our multi-shot implementation depends on the number of changepoints. 
In a practical two-day scenario, the number of changepoints typically falls within the range of fifteen to twenty-five, as highlighted in the grey area on the graph. In this range, the performance is deemed acceptable. Note also that even for sixty changepoints, the run-time remains well under that of the single-shot implementation for all but the coarsest granularities of times. We also implemented scenarios stretching over a week using our multi-shot implementation. Although there is a small increase in run-time compared to a two-day scenario with the same number of changepoints (probably due to the grounding size of the static part of the code), we can still conclude that the run-time indeed depends on the number of changepoints instead of the total number of timepoints.
On both graphs, namely Figure \ref{results} and Figure \ref{fig:changepointresults}, the average computation time per scenario is displayed based on 5 consecutive runs conducted on an Intel i5-8265U CPU. A repository containing all used scenarios is available online\footnote{\url{https://gitlab.com/EAVISE/bca/asp-for-flexible-payroll-management}}.

To use the system in practice, the continuous timeline needs to be split up into a number of independent scenarios. Typically this can be done by a single rule, e.g. in our example the end of a scenario coincides with the end of the shift of an employee, happening when they are absent for more than 4 hours.

\begin{figure}
\begin{tikzpicture}[scale=0.8]
\begin{axis}[
    xlabel={Number of timepoints},
    ylabel={Run time (seconds)},
    legend pos=north west,
    ymin=-10, ymax=500
]
\addplot[color=blue,mark=square]
    coordinates {
    (96,4.797)(115,5.203)(144,18.141)(192,25.844)(288,82.109)(576,446.46)
    };
\addplot[color=red,mark=square]
    coordinates {
    (96,1.14)(115,1.4128)(144,1.786)(192,2.214)(288,2.899)(576,4.15)
    };\legend{single shot implementation, multi-shot implementation}
\end{axis}
\end{tikzpicture}
\caption{Computation time of single-shot and multi-shot implementation}
\label{results}
\end{figure}
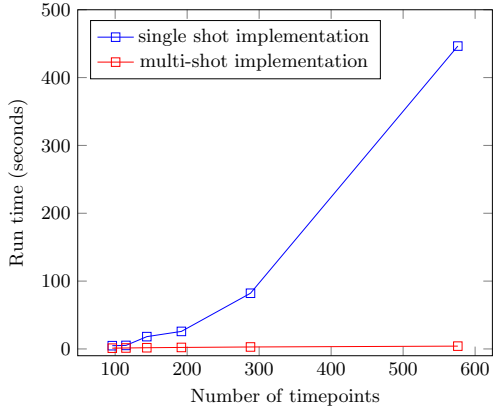
\begin{figure}
\begin{tikzpicture}[scale=0.8]
\begin{axis}[
    xlabel={Number of changepoints},
    ylabel={Run time (seconds)},
    ymin=-0.2, ymax=8, 
    legend pos=north west
]
\addplot[color=red,mark=square]
    coordinates {
    (60,3.7292)(55,3.1362)(50,2.7124)(45,2.4524)(40,2.3184)(35,2.1664)(30,1.6542)(25,1.4128)(20,1.2824)(15,0.8666)(10,0.508)
    };
\addplot[color=green,mark=square]
    coordinates {
    (60,7.368)(55,6.6638)(50,6.0718)(45,5.2532)(40,4.4376)(35,3.9530)(30,3.5312)
    };

\addplot[draw=none, fill=gray, fill opacity=0.5] coordinates {(15, 0) (15, 2.3) (25, 2.3) (25, 0)};

\legend{2-day scenarios, week scenarios}
\end{axis}
\end{tikzpicture}
\caption{Run time of multi-shot approach for various numbers of changepoints}
\label{fig:changepointresults}
\end{figure}
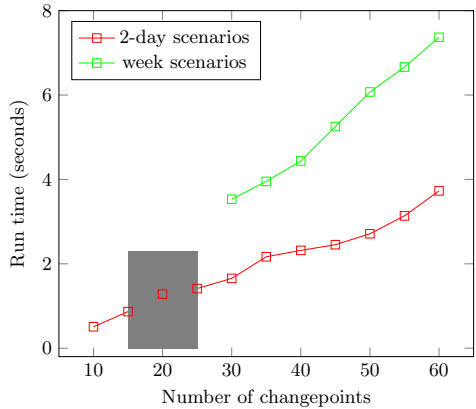

\section{Related Work}\label{relatedwork}

Different versions of the Event Calculus were formulated in ASP throughout the years. These implementations are both fast and expressive in comparison to previous SAT- or Prolog-based encodings, such as the DEC Reasoner introduced by \citeN{mueller2004tool} \citep{kim2009circumscriptive}. 
Given a finite domain, \textbf{EC2ASP} (and its evolution, F2LP) can compile both the EC and DEC formulas into ASP programs \citep{kim2009circumscriptive,lee2009system}. Our implementation of the DFEC axioms is closely related to their implementation of the DEC axioms. 

The \textbf{Functional Event Calculus} and its implementation to ASP were developped by \citeN{EFEC2014}. Their Epistemic Functional Event Calculus is able to deal with triggered, concurrent, non-deterministic, and conflicting action occurrences in a uniform manner under both discrete and continuous models of time. In this way, it can be used as a basis for a "possible-worlds" style approach to epistemic and causal reasoning in a narrative setting.
Recently, \textbf{s(CASP)} was also used to implement the Event Calculus formalism and reason about events without grounding the whole ASP program \citep{arias2019modeling}. s(CASP) is a top-down, goal-driven ASP system that can evaluate ASP programs with function symbols without grounding them \citep{arias2018constraint}. Due to the top-down nature of s(CASP), it is possible to elegantly model dense domains, such as time, as continuous quantities. In previous work \citep{lee2012reformulating,mellarkod2008integrating} such domains had to be discretized, thereby losing precision or even soundness. s(CASP) is thus able to faithfully model a continuous Event Calculus. The translation of EC axioms into a s(CASP) program is similar to that of the systems EC2ASP \citep{lee2012reformulating} but differs in some key aspects that improve performance and are relevant for expressiveness.  
However, since the task we try to solve in this paper explicitly requires us to consider the state of the employee at each timepoint, we would not benefit from the top-down approach of s(CASP). For similar reasons, a solution based on an Event Calculus implementation in Prolog, as done by \citeN{arias2019description}, would not result in beneficiary results. 

As recommended by \citeN{kim2009circumscriptive}, we use the clingo solver for our EC implementation. In addition to being an efficient single-shot solver, clingo also allows us to make use of a multi-shot solving strategy.
Next to clingo, DLV also developed an ASP solver with multi-shot reasoning capabilities: the \textbf{Incremental-DLV2 system} by \citeN{calimeri2022asp}. The system is built upon a proper integration of the overgrounding-based $I^{2}$-DLV incremental grounder \citep{ianni2020incremental} into DLV2, which employs the propositional WASP solver \citep{WASP}. In comparison to clingo, the approach of the Incremental-DLV2 is oriented towards declarative and transparent usage. Incremental-DLV2 implements overgrounding with tailoring \citep{ianni2020incremental}, an incremental grounding technique which, without requiring any operational statements from the user, incrementally drives the computation. In an XML file, the users can specify which parts of the logic program should be consecutively taken into account. Clingo, on the other hand, allows to procedurally control not only which parts of the logic program have to be taken into account during consecutive shots, but also how. In this way, clingo offers greater flexibility but also requires specific knowledge about how the system internally performs its computation and how the domain at hand is structured. Because these flexible control capabilities are needed to implement our multi-shot algorithm, we could only use clingo.

To our knowledge, there are no Event Calculus implementations in ASP that use the multi-shot solving capabilities in a similar way to efficiently reason about events and their effects.

\section{Conclusion}\label{conclusion}

In this paper, we have presented an approach to payroll management. We identified three key challenges for a payroll management system. HR consultants should be able to configure as much of the system as possible and the language in which they write down the rules must not only be easy to use for them, but it should also be possible to easily extend it with new language features. Finally, despite the required flexibility, the solution should still be computationally efficient. To tackle these challenges, we have split up the model into a generic Event-Calculus-based ASP program, and a decision-table-based model of the specific rules that apply in one particular set of circumstances. For the generic ASP program, we introduced a variant of the discrete Event Calculus that allows for non-boolean fluents: the Discrete Functional Event Calculus. This approach is not restricted to payroll management: its applicability can be extended to various domains, including planning and the monitoring of administrative processes, among other potential applications.

When used with a standard single-shot ASP solver, our implementation does not meet the computational requirements, due to the large number of timepoints that must be considered. We also present a multi-shot approach that eliminates this dependency on the absolute number of timepoints, by only considering those timepoints at which the state of the world changes. This multi-shot approach does reach the required performance.

In our future work,  we aim to implement payroll rules from a more diverse set of companies, spanning a wider range of countries and sectors to ensure comprehensive coverage. Additionally, we propose a formal validation of the complexity involved in defining payroll rules using our DMN tables, specifically targeting HR experts. Lastly, we intend to implement payroll rules that align with a company's legal obligations, including collective labor agreements. This approach would allow us to assess a company's compliance with legal standards.

\bibliographystyle{acmtrans.bst}
\bibliography{paper}

\appendix
\section{ASP implementation of DFEC}\label{appendix:DFEC}
\begin{lstlisting}[basicstyle=\scriptsize,escapechar=|,firstnumber=1]
% --------------------- List of domain-specific predicates ---------------------
% fluent(F). possVal(F,V). action(A). releases(A,F). causesValue(A,F,V,T).
% trajectory(F1,V1,T1,F2,V2,T2). antitrajectory(F1,V1,T1,F2,V2,T2).
% ----------------------- Domain independent DFEC axioms -----------------------
#const maxtime = 100.
time(0..maxtime).
% DFEC 1
valueCaused(F,V,T):-action(A),happens(A,T),causesValue(A,F,V),time(T).
% DFEC 2
initiated(F,V,T):-fluent(F),valueCaused(F,V,T),valueOf(F,V2,T),V2 != V.
% DFEC 3
terminated(F,V,T):-fluent(F),valueCaused(F,V2,T),valueOf(F,V,T),V2 != V.
% DFEC 4
stoppedIn(T1,F,V,T2):-fluent(F),time(T1),time(T2),time(T),
    terminated(F,V,T),T1 < T < T2.
% DFEC 5
startedIn(T1,F,V,T2):-fluent(F),time(T1),time(T2),time(T),
    initiated(F,V,T),T1 < T < T2.
% DFEC 6
valueOf(F2,V2,T1+T2):-fluent(F1),time(T1),initiated(F1,V1,T1),fluent(F2),
    time(T2),trajectory(F1,V1,T1,F2,V2,T2),not stopppedIn(T1,F1,V1,T1+T2).
% DFEC 7
valueOf(F2,V2,T1+T2):-fluent(F1),time(T1),terminated(F1,V1,T1),fluent(F2),
    time(T2),antitrajectory(F1,V1,T1,F2,V2,T2),not startedIn(T1,F1,V1,T1+T2).
% DFEC 8
valueOf(F,V,T+1):-fluent(F),time(T+1),valueOf(F,V,T),not terminated(F,V,T),
    not releasedAt(F,T).
% DFEC 9
releasedAt(F,T+1):-fluent(F),time(T),time(T+1),releasedAt(F, T),possVal(F,V),
    not initiated(F,V,T),not terminated(F,V,T).
% DFEC 10
released(F,T):-fluent(F),time(T),releases(A, F),action(A),happens(A,T).
:- not releasedAt(F,T),not released(F,T),releasedAt(F,T+1),time(T+1), 
    time(T),fluent(F).
% DFEC 11
valueOf(F,V,T+1):-fluent(F),time(T+1),initiated(F,V,T).
% DFEC 12
:- terminated(F,V,T),valueOf(F,V,T+1),fluent(F),time(T+1).
% DFEC 13
releasedAt(F,T+1):-fluent(F),time(T+1),action(A),releases(A,F), happens(A,T).
% DFEC 14
:- initiated(F,V,T),releasedAt(F,T+1),T < maxtime.
:- terminated(F,V,T),releasedAt(F,T+1),T < maxtime.
% DFEC15
:- valueOf(F,V,T),not possVal(F,V).
\end{lstlisting}
\section{Proof of stratification DFEC implementation}\label{appendix:Mapping}
To prove that DFEC implementation in ASP within a finite domain is stratified, a level mapping of the DFEC implementation is displayed in Table \ref{tab:level_mapping}. Note that for all predicates, $0<I$ holds. For $stoppedIn(J,F,V,I)$ and $startedIn(J,F,V,I)$, $0<J<I$ holds.

\begin{table}[h]
    \centering
    \footnotesize
    \begin{tabular}{|c|c|c|c|c|}
        \hline
        \multicolumn{2}{|c|}{\textbf{Level 0}} & \textbf{Level I} & \textbf{Level I+1}\\
        \hline
        \cline{1-2}
        fluent(F) & time(T) & stoppedIn(J,F,V,I) & valueOf(F,V,I)\\
        valueCaused(F,V,T) & possVal(F,V) & startedIn(J,F,V,I) & initiated(F,V,I)\\
        action(A) & happens(A,T) &  & terminated(F,V,I)\\
        causesValue(A,F,T) & releases(A,F)  &  & releasedAt(F,I)\\
        released(F,T)&&&\\
        \multicolumn{2}{|c|}{trajectory(F1,V1,T1,F2,V2,T2)} & &\\
        \multicolumn{2}{|c|}{antitracetory(F1,V1,T1,F2,V2,T2)} & &\\
        \hline
    \end{tabular}
    \caption{Level mapping of DFEC implementation}
    \label{tab:level_mapping}
\end{table}
\vspace{-1.5em}
\section{General ASP multi-shot implementation}\label{appendix:MS}
\begin{lstlisting}[basicstyle=\scriptsize,firstnumber=1]
#program static().
% --- Company Specific Rules ---
% fluent(F) boolean(F) domain(F,V)
% userAction(A) walltimAction(A) triggeredAction(A) actionTime(A,T)
% causes(A, F, V) ccauses(A,F,V,Cond)
% and(Cond,N) sub(Cond,I,SubCond)
% after(F, V, T, A) when(CF,T,V)
% relevant(F)
% defined(F) rule(F,V,Cond)
% countFluent(F) countRule(CF,F,V).
% --- User Specific Scenario ---
% userDoes(clockIn,T) userDoes(clockOut,T)
% --- Domain independent Axioms ---
day(0..1).  hour(0..23).   minute(0..59).
maxTime(Max):- Max =  #max{Stamp: ts(Days,Hours,Minutes,Stamp)}.
ts(Days, Hours, Minutes, Stamp) :- day(Days), hour(Hours), 
    minute(Minutes),Stamp = (Days*24+Hours)*60+Minutes.
domain(F, (true;false)) :- boolean(F).
initially(F,false) :- fluent(F), boolean(F).
action(A) :- userAction(A).
action(A) :- walltimeAction(A).
action(A) :- triggeredAction(A).
%Timepoint 0
timepoint(0).
value(CF,0,0) :- countFluent(CF).
value(F,Val,0) :- fluent(F), initially(F,Val).
initiated(F,Val,0) :- fluent(F), value(F,Val,0).
default(V,false) :- defined(V), boolean(V).
value(V,Val,0) :- defined(V),default(V,Val).
boundary(0).

#program dynamic(pp,cp).
% --- Domain independent Axioms ---
next(pp,cp).
timepoint(cp).
initiated(F,Val,cp) :- fluent(F),cause(F,Val,cp),value(F,V2,cp), 
    V2 != Val.
terminated(F,V2,cp) :- fluent(F),cause(F,Val,cp),value(F,V2,cp), 
    V2 != Val.
value(F,Val,cp) :- fluent(F), timepoint(cp), value(F,Val,pp), 
    not terminated(F,Val,pp).
value(F,Val,cp) :- fluent(F), timepoint(cp), initiated(F,Val,pp).
value(CF,S+S1,cp) :- countFluent(CF),timepoint(pp),value(CF,S,pp),
    countRule(CF,F,V),value(F,V,cp), S1=cp-pp.
value(CF,S,cp) :- countFluent(CF),timepoint(pp),value(CF,S,pp),
    countRule(CF,F,V),value(F,V1,cp), V != V1.
value(Var, Val, cp) :- defined(Var), timepoint(cp), 
    rule(Var,Val,Cond), holds(Cond,cp).
value(Var, Val, cp) :- defined(Var), timepoint(cp), 
    default(Var,Val), not applicableRule(Var,cp).
applicableRule(Var,cp) :- defined(Var), rule(Var, Val, Cond), 
    holds(Cond,cp).
:- value(F,V1,cp),value(F,V2,cp),V1!=V2.
cause(F,Val,cp) :- action(A), happens(A,cp), 
    causes(A,F,Val,cp), timepoint(cp).
causes(A,F,Val,cp) :- causes(A,F,Val),timepoint(cp).
causes(A,F,Val,cp) :- ccauses(A,F,Val,Cond),holds(Cond,cp).
holds(v(F,Val),cp) :- value(F,Val,cp), timepoint(cp).
holds(sinceBetween(F,Val,Min,Max),cp) :- timepoint(cp),
    sameSince(F,Val,T1,cp),initiated(F,Val,T1),
    timepoint(T1),pbreak(Min),pbreak(Max),Min<=cp-T1,cp-T1<Max.
holds(sinceMin(F,Val,Min),cp) :-timepoint(cp),sameSince(F,Val,T1,cp),
    initiated(F,Val,T1),timepoint(T1),pbreak(Min),Min<=cp-T1.
holds(C,cp) :- and(C,Nb), timepoint(cp), #count{I : sub(C,I,Sub),
    holds(Sub,cp)} = Nb.
happens(A,cp) :- userAction(A),userDoes(A,cp),timepoint(cp).
happens(A,cp) :- walltimeAction(A),actionTime(A,cp),timepoint(cp).
happens(A,cp) :- after(F,V,D,A),timepoint(T1),timepoint(cp),
    cp=T1+D,sameSince(F,V,T1,cp),initiated(F,V,T1).
happens(A,cp) :- timepoint(cp),when(CF,V,A),value(CF,V,cp),
    value(CF,V1,pp),V != V1.
changed(Var,Old,pp):-value(Var,Old,pp),value(Var,New,cp),Old!=New.
changesTo(Var,New,pp):-value(Var,Old,pp),value(Var,New,cp),Old!=New.
sameSince(F,V,T1,cp):-timepoint(pp),sameSince(F,V,T1,pp),
    not changed(F,V,pp).
sameSince(F,V,pp,cp):-changesTo(F,V,pp).
boundary(pp) :- relevant(Var),changed(Var,Val,pp).

#program boundaries(i,start,end).
% --- Domain independent Axioms ---
intervalFrom(i,start).
intervalTo(i,end).

#program interval().
% --- Company rule-specific axioms ---
% defined(F). domain(F,V).
% rule(F,V,Cond) iand(Cond,N) isub(Cond,X,SC).
boundary(I,T):-intervalFrom(I,T).
boundary(I,T):-intervalTo(I,T).
id(0..20).
enoughIds :- freeId(I).
freeId(I) :- id(I), maxId(M), I > M.
maxId(M) :- id(M), M =  #max{I: usedId(I) }.
usedId(I) :- id(I), boundary(I,T).
:- not enoughIds.
length(Int, L):-intervalFrom(Int, From),intervalTo(Int,To),L=To-From.
shorter(Id1,Id2) :- length(Id1,L1),length(Id2,L2),L1<L2.
intHolds(v(F,Val), Int):-value(F,Val,T), domain(F,Val),
    id(Int), intervalTo(Int,T), timepoint(T).
intHolds(C,Int) :- iand(C,Nb), id(Int), 
    #count {I : isub(C,I,Sub), intHolds(Sub,Int)} = Nb.
intHolds(shorter(I,J),Int):-id(Int),id(Int1),id(Int2),
    refersTo(I,Int1,Int),refersTo(J,Int2,Int),shorter(Int1,Int2).
intHolds(otherv(Fluent,Val,J),Int):-id(Int),id(Int1),
    refersTo(J,Int1,Int),intHolds(v(Fluent,Val),Int1).
refersTo(this,Int1,Int) :- id(Int),id(Int1),Int=Int1.
refersTo(next(I), J+1, Int) :- id(Int),id(J),refersTo(I,J,Int).
intVal(F,V,Int):-defined(F),id(Int),value(F,V,T),intervalTo(Int,T).
workedHours(S,Int):-value(workedHours,S,T),intervalTo(Int,T).
intValue(Var,Val,Int):-defined(Var),id(I),rule(Var,Val,Cond),
    intHolds(Cond,I).
interval(I,From,To):-id(I),intervalFrom(I,From),intervalTo(I,To).
\end{lstlisting}
\end{document}